\documentclass[11pt]{article}
\usepackage{amssymb,amsmath,mathrsfs,enumerate}
\usepackage{graphicx,rotate,multicol}
\usepackage[margin=10pt,labelfont=bf]{caption}
\usepackage{cite}
\usepackage{braket}
\usepackage[utf8]{inputenc}
\usepackage[colorlinks=true,
            linkcolor=red,
            urlcolor=blue,
        citecolor=blue]{hyperref}

\usepackage{color}
\usepackage{lineno}
\long\def\rpl#1!!!#2!!!{\color[rgb]{.7,0,0}{#1} \color{blue}{#2} \color{black}}
\let\tilde=\widetilde

\let\bar=\overline

\def \order(#1){{\mathcal O} \left(#1 \right)}
\def\rep#1{\ensuremath{\mathbf #1}}

\evensidemargin=\oddsidemargin
\def\Eqn#1{Eq.\ (\ref{#1})}
\def\Eqs#1#2{Eqs.\ (\ref{#1}) and (\ref{#2})}
\allowdisplaybreaks

\date{}

\begin{document}
\begin{flushright}
	LU TP 19-39
\end{flushright}	
%\pagewiselinenumbers

\begin{center}
	{\Large \bf Relating the Cabibbo angle to $\mathbb{\tan\beta}$ in a
		two Higgs-doublet model} \\
	\vspace*{0.5cm} {\sf Dipankar Das\footnote[1]{dipankar.das@thep.lu.se}} \\
	\vspace{10pt} {\small \em 
		%%%%%%%%%%%%%%%%%%%%%%%%%%%%%%%%%%%%%%%%%%%%%%%%%%%%%%%%%%%%%
		Department of Astronomy and Theoretical Physics, Lund University, S\"{o}lvegatan 14A, Lund 22362, Sweden}
	
	\normalsize
\end{center}

%\renewcommand*{\thefootnote}{\arabic{footnote}}
%\setcounter{footnote}{0} 

%\maketitle

\begin{abstract}
In a two Higgs-doublet model with $D_4$ flavor symmetry we
establish a relation between $\tan\beta$ and the Cabibbo
angle. Due to small number of parameters, the quark Yukawa
sector of the model is very predictive. The flavor changing
neutral currents are small enough to allow for relatively
light nonstandard scalars to pass through the flavor constraints.
\end{abstract}

\bigskip
%%%%%%%%%%%%%%%%%%%%%%%%%%%%%%%%%%%%%%%%%%%%%%%%%%%%%%
%%%%%%%%%%   Main Text Starts here   %%%%%%%%%%%%%%%%%
%%%%%%%%%%%%%%%%%%%%%%%%%%%%%%%%%%%%%%%%%%%%%%%%%%%%%%

Employing flavor symmetries to understand the apparent
arbitrariness of the quark masses and mixings in the Standard
Model~(SM) is an exercise continuing for decades. The Yukawa
Lagrangian of the SM also contains many redundant parameters,
which is not a very attractive feature of the model. Therefore,
theoretical constructions beyond the SM~(BSM) that attempt to
address these issues in a minimalistic manner should deserve
some attention. To this end, we notice that the quark masses
and mixings adhere to the following approximate pattern,
\begin{eqnarray}
\label{e:approx}
m_u \approx 0 \,, \quad m_d \approx 0 \,, \quad V_{\rm CKM} =
\begin{pmatrix}
\cos\theta_C & \sin\theta_C & 0 \\
-\sin\theta_C & \cos\theta_C & 0 \\
0 & 0 & 1
\end{pmatrix},
\end{eqnarray}
where $V_{\rm CKM}$ stands for the Cabibbo–Kobayashi–Maskawa (CKM) matrix
with only the Cabibbo block retained. The quantity 
$\sin\theta_C\approx 0.22$ appearing in \Eqn{e:approx} denote
the Cabibbo mixing parameter. In this approximate scenario, we
note that there are only five nonzero parameters in the quark
sector, namely, four quark masses ($m_c$, $m_s$, $m_t$, $m_b$)
and the Cabibbo parameter itself. Thus, a flavor-model that
contains five or fewer parameters in its quark Yukawa Lagrangian,
might have a better aesthetic appeal than the SM in the sense
that many of the redundant parameters have been erased by the
flavor symmetry leaving behind only the relevant ones. The
model can be even more attractive, if the zeros in \Eqn{e:approx}
emerge naturally as a consequence of the Yukawa textures
imposed by the flavor symmetry. As we will demonstrate, these
objectives can be achieved in the simple framework of a two
Higgs-doublet model (2HDM)\cite{Branco:2011iw,Bhattacharyya:2015nca} 
with a $D_4$ flavor symmetry.

The discrete symmetry group $D_4$ has five irreducible
representations: \rep{1_{++}}, \rep{1_{+-}}, \rep{1_{-+}},
\rep{1_{--}} and \rep{2}\cite{Ishimori:2010au,pal_2019}.  We pick a basis such that
the generators in the \rep{2} representation are given by
\begin{eqnarray}
a = \begin{bmatrix} 0 & -1 \\ 1 & 0 \end{bmatrix} \,, 
\qquad
b = \begin{bmatrix} 1 & 0  \\ 0 & -1 \end{bmatrix}\,.
\label{gen}
\end{eqnarray}
Note that $a$ is of order 4, whereas $b$ is of order 2.  The rest of
the elements can be obtained by taking products of powers of these two
elements. In this basis, the relevant tensor products are obtained
as
\begin{subequations}
\label{e:tp}
\begin{eqnarray}
\begin{bmatrix} x_1 \\ x_2 \end{bmatrix}_{\rep{2}} \otimes 
\begin{bmatrix} y_1 \\ y_2 \end{bmatrix}_{\rep{2}} &=& 
[x_1y_1+x_2y_2]_{\rep{1_{++}}} \oplus [x_1y_2-x_2y_1]_{\rep{1_{--}}} \nonumber \\
&& \oplus \, [x_1y_2+x_2y_1]_{\rep{1_{-+}}} \oplus [x_1y_1-x_2y_2]_{\rep{1_{+-}}} \,, \\
\rep{1_{rs}} \otimes \rep{1_{r's'}} &=& \rep{1_{r{''}s{''}}} \,,
\end{eqnarray}
\end{subequations}
where $r{''}=r\cdot r'$ and $s{''}=s\cdot s'$.
The quark fields are assumed to transform under $D_4$ in
the following way:
\begin{subequations}
	\label{quarkD4}
	\begin{eqnarray}
	&& {\bf 2}  \,\, : \,\, \begin{bmatrix}Q_1 \\ Q_2 \end{bmatrix},
	\; \begin{bmatrix}p_{1R} \\ p_{2R} \end{bmatrix},
	\; \begin{bmatrix}n_{1R} \\ n_{2R} \end{bmatrix}\,, 
	\label{qLS3} \\*
	&& \rep{1_{++}} \,\, : \,\, Q_3 \,, \qquad 
	\rep{1_{--}} \,\, : \,\, p_{3R} \,, \qquad \rep{1_{-+}} \,\, : \,\, n_{3R} \,,
	\label{qRS3}
	\end{eqnarray}
\end{subequations}
where the $Q_A$'s ($A=1,2,3$) are the usual left-handed $\rm SU(2)$
quark doublets, whereas the $p_{AR}$'s and $n_{AR}$'s are the
right-handed up-type and down-type quark fields respectively, which
are singlets of the $\rm SU(2)$ part of the gauge symmetry.  Note that
the square brackets in Eqs.~(\ref{gen}), (\ref{e:tp}) and (\ref{quarkD4}) as well as in the
subsequent text, denote the representations of $D_4$ and has
nothing to do with the representation of the enclosed fields under
$\rm SU(2)$. In the Higgs sector there are two $\rm SU(2)$
doublets $\phi_k$ ($k=1,2)$ and their transformation under the $D_4$
symmetry is as follows:
\begin{eqnarray}
\rep{2} & : & \begin{bmatrix} \phi_1 \\ \phi_2 \end{bmatrix} \,.  
%\label{}
\end{eqnarray}
The most general Yukawa Lagrangian for the
quarks that is consistent with the gauge and $D_4$ symmetries
can be written as
\begin{eqnarray}
 - {\mathscr L}_Y &=&
\null A_u \Big( 
\bar Q_1 \tilde\phi_2 - \bar Q_2\tilde\phi_1 \Big)
p_{3R} + B_u \, \bar Q_3 \Big( \tilde\phi_1 p_{1R} +
\tilde\phi_2 p_{2R} \Big) \nonumber \\
&& + A_d \Big( 
\bar Q_1 \phi_2 + \bar Q_2 \phi_1 \Big)
n_{3R} + B_d \, \bar Q_3 \Big(\phi_1 n_{1R} +
\phi_2 n_{2R} \Big)
+ {\rm h.c.} \,,
\label{e:Yuk}
\end{eqnarray}
where, we have used the standard abbreviation $\tilde\phi_k = i \sigma_2
\phi_k^*$. The complex phases of the Yukawa couplings can be absorbed in
the quark field redefinitions.
Thus, the $D_4$ symmetry reduces the number of Yukawa couplings
drastically to the extent that we are left with only five unknown
parameters in \Eqn{e:Yuk}, namely, four Yukawa couplings and the ratio
of the two vacuum expectation values (VEVs), $\tan\beta \equiv
v_2/v_1$. Quite remarkably, these are just enough to reproduce the 
five nonzero parameters in the quark sector when \Eqn{e:approx} 
holds. Therefore, at this leading order, using a $D_4$ flavor symmetry
we have successfully removed all the unnecessary parameters from the
quark Yukawa Lagrangian.
The mass matrices that follow from \Eqn{e:Yuk} are given by
\begin{eqnarray}
M_u = \frac{1}{\sqrt{2}} \begin{pmatrix} 0 & 0 & A_u v_2 \\ 0 & 0 & -A_u v_1 \\
B_u v_1 & B_u v_2 & 0 \end{pmatrix} , \qquad
M_d = \frac{1}{\sqrt{2}} \begin{pmatrix} 0 & 0 & A_d v_2 \\ 0 & 0 & A_d v_1 \\
B_d v_1 & B_d v_2 & 0 \end{pmatrix} ,  
\label{e:mumd}
\end{eqnarray}
where $\braket{\phi_k}=v_k/\sqrt{2}$ represents the VEV of $\phi_k$.
The diagonal mass matrices can be obtained via the following biunitary transformations:
\begin{subequations}
	\label{e:bidiag}
	\begin{eqnarray}
	 && D_u = V_L\cdot M_u \cdot V_R^\dagger = {\rm diag}(m_u,m_c,m_t) \,, \\
	 && D_d = U_L\cdot M_d \cdot U_R^\dagger = {\rm diag}(m_d,m_s,m_b) \,.
	\end{eqnarray}
\end{subequations}
The matrices, $V$ and $U$ relate the quark fields in the gauge basis to 
those in the mass basis as follows:
\begin{subequations}
	\label{e:ud}
	\begin{eqnarray}
	 u_L = V_L \, p_L \,, && u_R = V_R \, p_R \,, \\
	 d_L = U_L\, n_L \,, && d_R = U_R \, n_R \,,
	\end{eqnarray}
\end{subequations}
where, $u$ and $d$ denote the physical up and down type
quarks respectively.
The CKM matrix is then given by
\begin{eqnarray}
\label{e:CKMdef}
	V_{\rm CKM} = V_L \cdot U_L^\dagger \,.
\end{eqnarray}
The matrices, $V_L$ and $U_L$ can be obtained by diagonalizing $M_u M_u^\dagger$
and $M_d M_d^\dagger$ respectively, which can be calculated from \Eqn{e:mumd}
as follows:
\begin{eqnarray}
M_u M_u^\dagger= \frac{1}{2} \begin{pmatrix} A_u^2 v_2^2 & -A_u^2 v_1v_2 & 0 \\ 
-A_u^2 v_1v_2 & A_u^2 v_1^2 & 0 \\ 0 & 0 & B_u^2 v^2 \end{pmatrix} , \qquad
M_d M_d^\dagger = \frac{1}{2} \begin{pmatrix} A_d^2 v_2^2 & A_d^2 v_1v_2 & 0 \\ 
A_d^2 v_1v_2 & A_d^2 v_1^2 & 0 \\ 0 & 0 & B_d^2 v^2 \end{pmatrix} ,  
\label{e:ulvl}
\end{eqnarray}
where, $v=\sqrt{v_1^2+v_2^2}$ is the total electroweak VEV.
To diagonalize the above matrices, we introduce the matrix,
\begin{eqnarray}
U_\beta = \begin{pmatrix} \cos\beta & \sin\beta & 0 \\ -\sin\beta & \cos\beta & 0 \\
0 & 0 & 1 \end{pmatrix}.  
\label{e:ub}
\end{eqnarray}
One can easily check that
\begin{subequations}
	\label{e:diag}
	\begin{eqnarray}
	&& D_u^2 = U_\beta \cdot (M_uM_u^\dagger) \cdot U_\beta^\dagger = 
	{\rm diag}\left(0,A_u^2 v^2/2,B_u^2 v^2/2\right) \,, \\
	&& D_d^2 = U_\beta^\dagger \cdot (M_dM_d^\dagger) \cdot U_\beta = 
	{\rm diag}\left(0,A_d^2 v^2/2,B_d^2 v^2/2\right) \,.
	\end{eqnarray}
\end{subequations}
Thus, we can identify the masses of the physical quarks as
\begin{eqnarray}
m_{u,d}^2 = 0 \,, \qquad  m_{c,s}^2 = \frac{1}{2} A_{u,d}^2 v^2 \,,
\qquad  m_{t,b}^2 = \frac{1}{2} B_{u,d}^2 v^2 \,.
\label{e:masses}
\end{eqnarray}
Also, comparing with the definitions in \Eqn{e:bidiag}, we can conclude
\begin{eqnarray}
V_L = U_\beta \,, \qquad  U_L = U_\beta^\dagger \,.
\label{e:uldl}
\end{eqnarray}
Using \Eqn{e:CKMdef} we can now easily calculate the CKM matrix as
follows:
\begin{eqnarray}
\label{e:CKM}
V_{\rm CKM} = U_\beta \cdot U_\beta = \begin{pmatrix} \cos 2\beta & \sin 2\beta & 0 \\
 -\sin 2\beta & \cos 2\beta & 0 \\ 0 & 0 & 1 \end{pmatrix}.
\end{eqnarray}
Therefore, comparing with \Eqn{e:approx}, one can identify the Cabibbo mixing angle as
\begin{eqnarray}
\label{e:key}
\sin\theta_C = \sin 2\beta \approx 0.22 \,.
\end{eqnarray}
This relation between the Cabibbo parameter and $\tan\beta$ is the key result of our analysis. Note that the relation of \Eqn{e:key} is
purely a consequence of the Yukawa textures in \Eqn{e:mumd} which are
dictated by the $D_4$ symmetry. Therefore, this relation should be
stable under quantum corrections.

%%%%%%%%%%%%%%%%%%%%%%%%%%%%%%%%%%%%%%%%%%%%%%%%%%%%%%%%%%%%%%%
%%%%%%%%%%%%%%  The Scalar Sector  %%%%%%%%%%%%%%%%%%%%%%%%%
%%%%%%%%%%%%%%%%%%%%%%%%%%%%%%%%%%%%%%%%%%%%%%%%%%%%%%%%%%%%
At this stage, it is reasonable to ask whether such a value of $\tan\beta$
will be allowed from the scalar sector. As we will see, the value of $\tan\beta$
can be quite arbitrary if we allow for terms that softly break the $D_4$ symmetry
in the scalar sector. Keeping these in mind, we write the scalar potential
as
\begin{eqnarray}
V &=& -\mu_1^2 \left(\phi_1^\dagger \phi_1\right) -\mu_2^2 \left(\phi_2^\dagger \phi_2\right)
-\mu_{12}^2 \left(\phi_1^\dagger \phi_2 + \phi_2^\dagger \phi_1 \right)
+ \lambda_1 \left(\phi_1^\dagger \phi_1 + \phi_2^\dagger \phi_2 \right)^2 \nonumber \\
&& + \lambda_2 \left(\phi_1^\dagger \phi_2 - \phi_2^\dagger \phi_1 \right)^2 
+ \lambda_3 \left(\phi_1^\dagger \phi_2 + \phi_2^\dagger \phi_1 \right)^2
+ \lambda_4 \left(\phi_1^\dagger \phi_1 - \phi_2^\dagger \phi_2 \right)^2 \,.
\label{e:pot}
\end{eqnarray}
Note that, in the limit $\mu_1^2=\mu_2^2$, $\mu_{12}^2=0$ the $D_4$ symmetry will
be exact in the scalar potential. However, in this case one can easily verify that
the minimization conditions will enforce $v_1=v_2$, {\it i.e.}, $\tan\beta=1$
which will be incompatible with \Eqn{e:key}. Therefore, we decide to proceed with
the potential of \Eqn{e:pot} containing the most general bilinear terms. The
minimization conditions, in this case, can be used to solve for the bilinear
parameters $\mu_1^2$ and $\mu_2^2$ as follows:
\begin{subequations}
\label{e:min}
\begin{eqnarray}
\mu_1^2 &=& (\lambda_1 +2\lambda_3-\lambda_4) v_2^2 + (\lambda_1+\lambda_4) v_1^2
+\mu_{12}^2 \frac{v_2}{v_1} \,, \\
\mu_2^2 &=& (\lambda_1 +2\lambda_3-\lambda_4) v_1^2 + (\lambda_1+\lambda_4) v_2^2
+\mu_{12}^2 \frac{v_1}{v_2} \,.
\end{eqnarray}
\end{subequations}
After the spontaneous symmetry breaking, we expand the scalar doublets
as
\begin{eqnarray}
	\phi_k =\frac{1}{\sqrt{2}} \begin{pmatrix} \sqrt{2} w_k^+ \\ v_k+h_k+iz_k
	\end{pmatrix} \qquad (k=1,2) \,.
\end{eqnarray}
The massless unphysical scalars $\omega^\pm$ and $\zeta$ in the charged and 
the pseudoscalar sectors respectively, can be extracted using the following rotation:
\begin{eqnarray}
\begin{pmatrix} \omega^\pm \\ H^\pm \end{pmatrix} =
\begin{pmatrix} \cos\beta & \sin\beta \\ -\sin\beta & \cos\beta \end{pmatrix}
\begin{pmatrix} w_1^\pm \\ w_2^\pm \end{pmatrix} , \qquad
\begin{pmatrix} \zeta \\ A \end{pmatrix} =
\begin{pmatrix} \cos\beta & \sin\beta \\ -\sin\beta & \cos\beta \end{pmatrix}
\begin{pmatrix} z_1 \\ z_2 \end{pmatrix} .
\end{eqnarray}
In the above equation, $H^\pm$ and $A$ stand for physical charged scalar and
pseudoscalar respectively, whose masses can be calculated as
\begin{subequations}
\label{e:mpma}
\begin{eqnarray}
m_+^2 &=& \frac{2\mu_{12}^2}{\sin 2\beta} -2\lambda_3 v^2 \,, \\
m_A^2 &=& \frac{2\mu_{12}^2}{\sin 2\beta} -2(\lambda_2+\lambda_3) v^2 \,.
\end{eqnarray}
\end{subequations}
The mass squared matrix in the scalar sector is given by
\begin{subequations}
\label{e:CPe}
\begin{eqnarray}
&& V_S^{\rm mass} = \begin{pmatrix}
h_1 & h_2 \end{pmatrix} \frac{{\cal M}_S^2}{2}
\begin{pmatrix}  h_1 \\ h_2 \end{pmatrix} , \\
{\rm with,} &&
{\cal M}_S^2 = \begin{pmatrix}
2(\lambda_1+\lambda_4)v_1^2 +\mu_{12}^2 \frac{v_2}{v_1} &
2(\lambda_1+2\lambda_3 -\lambda_4)v_1 v_2 - \mu_{12}^2 \\
2(\lambda_1+2\lambda_3 -\lambda_4)v_1 v_2 - \mu_{12}^2 &
2(\lambda_1+\lambda_4)v_2^2 +\mu_{12}^2 \frac{v_1}{v_2}
\end{pmatrix}.
\end{eqnarray}
\end{subequations}
The diagonalization of ${\cal M}_S^2$ will lead to two
physical $CP$-even scalars $H$ and $h$ which are obtained
via the following rotation
\begin{eqnarray}
    \begin{pmatrix} H \\ h \end{pmatrix} =
    \begin{pmatrix} \cos\alpha & \sin\alpha \\ 
    -\sin\alpha & \cos\alpha \end{pmatrix}
    \begin{pmatrix} h_1 \\ h_2 \end{pmatrix}.
\end{eqnarray}
This diagonalization will then entail the following relations:
\begin{subequations}
\label{e:scalar}
\begin{eqnarray}
m_H^2\cos^2\alpha +m_h^2\sin^2\alpha &=&
2(\lambda_1+\lambda_4)v_1^2 +\mu_{12}^2 \frac{v_2}{v_1} \,, \\
m_H^2\sin^2\alpha +m_h^2\cos^2\alpha &=&
2(\lambda_1+\lambda_4)v_2^2 +\mu_{12}^2 \frac{v_1}{v_2} \,, \\
(m_H^2-m_h^2)\sin\alpha \cos\alpha &=&
2(\lambda_1+2\lambda_3 -\lambda_4)v_1 v_2 - \mu_{12}^2 \,.
\end{eqnarray}
\end{subequations}
We note that the potential of \Eqn{e:pot} contains seven
parameters among which two of the bilinear parameters, $\mu_1^2$
and $\mu_2^2$, have been traded in favor of $v_1$ and $v_2$ (or
equivalently $v$ and $\tan\beta$) using \Eqn{e:min}. The
remaining five parameters (four lambdas and $\mu_{12}^2$) can
then be exchanged for four physical masses ($m_+$, $m_A$, $m_H$
and $m_h$) and the mixing angle, $\alpha$ using 
\Eqs{e:mpma}{e:scalar}. On top of this,
putting $\alpha=\beta-\pi/2$\cite{Gunion:2002zf} will ensure that $h$ possesses
exact SM-like couplings at the tree-level, so that it can be
identified with the $125$~GeV scalar discovered at the LHC.
In this {\em alignment limit}\cite{Das:2015mwa,Dev:2014yca}, \Eqn{e:scalar} can be rearranged
to obtain simpler expressions for $m_h$ and $m_H$ as follows:
\begin{subequations}
\label{e:mhmH}
\begin{eqnarray}
m_h^2 &=& 2(\lambda_1+\lambda_3) v^2 \,, \\
m_H^2 &=& \frac{2\mu_{12}^2}{\sin 2\beta} +2(\lambda_4-\lambda_3) v^2 \,.
\end{eqnarray}
\end{subequations}
From \Eqs{e:mpma}{e:mhmH} we can see that, in the limit
$\mu_{12}^2 \gg v^2$, only the SM-like Higgs scalar, $h$,
remains at the EW scale while the other nonstandard scalars
are quasidegenerate and super heavy.
Considering the absence of any convincing hints of new physics
at the collider experiments, such a spectrum of the scalar masses
might be desirable. Moreover,
in the limit $m_+ \approx m_H \approx
m_A \gg m_h$, the bound from the electroweak $T$-parameter
can be easily avoided\cite{Grimus:2007if,Bhattacharyya:2013rya}.

For the sake of completeness we now discuss the scalar mediated
flavor changing neutral currents~(FCNCs) in our model.
Comparing \Eqn{e:Yuk} with the general 2HDM Yukawa Lagrangian
\begin{eqnarray}
{\mathscr L}_Y = -\sum_{k=1}^{2} \left[\bar{Q}\,\Gamma_k\,
\phi_k\, n_R +\bar{Q}\,\Delta_k\, \tilde{\phi_k}\, p_R \right]
+ {\rm h.c.}\,, \label{e:Yukgen}
\end{eqnarray}
we can write,
\begin{eqnarray}
\Delta_1 = \begin{pmatrix}
0 & 0 & 0 \\ 0 & 0 & -A_u \\ B_u & 0 & 0 \end{pmatrix},
\quad
\Delta_2 = \begin{pmatrix}
0 & 0 & A_u \\ 0 & 0 & 0 \\ 0 & B_u & 0 \end{pmatrix},
\quad
\Gamma_1 = \begin{pmatrix}
0 & 0 & 0 \\ 0 & 0 & A_d \\ B_d & 0 & 0 \end{pmatrix},
\quad
\Gamma_2 = \begin{pmatrix}
0 & 0 & A_d \\ 0 & 0 & 0 \\ 0 & B_d & 0 \end{pmatrix}.
\label{e:textures}
\end{eqnarray}
Note that in writing \Eqn{e:Yukgen}, we have suppressed
the generation indices. The matrices, $N_u$ and $N_d$,
which control the FCNC couplings in the up and down sectors
respectively, are given by\cite{Branco:2011iw}
\begin{subequations}
\label{e:nund}
\begin{eqnarray}
N_u &=& \frac{1}{\sqrt{2}} V_L \left(\Delta_1 v_2
-\Delta_2 v_1 \right) V_R^\dagger \,, \\
N_d &=& \frac{1}{\sqrt{2}} U_L \left(\Gamma_1 v_2
-\Gamma_2 v_1 \right) U_R^\dagger \,.
\end{eqnarray}
\end{subequations}
As an explicit example, $N_u$ and $N_d$ will get involved in
the FCNC couplings in the physical $CP$-even sector as follows
\begin{eqnarray}
{\mathscr L}_Y^{\rm CP~even} = -\frac{h}{v} \left(
\bar{u}\, D_u\, u +\bar{d}\, D_d\, d \right) -\frac{H}{v} \left[
\bar{u}\left(N_uP_R+N_u^\dagger P_L\right)u +\bar{d}
\left(N_dP_R+N_d^\dagger P_L \right) d\right] \,,
\end{eqnarray}
where, we have suppressed again the generation indices and
imposed the alignment limit. To calculate the expressions
for $N_u$ and $N_d$ using \Eqn{e:nund}, we need to know
$V_R$ and $U_R$ which can be obtained by diagonalizing $M_u^\dagger M_u$
and $M_d^\dagger M_d$ respectively. In this way, we find
\begin{eqnarray}
V_R = U_R = \begin{pmatrix}
-\sin\beta & \cos\beta & 0 \\ 0 & 0 & 1 \\
\cos\beta & \sin\beta & 0
\end{pmatrix} \,.
\end{eqnarray}
Now we can easily compute $N_u$ and $N_d$ as follows:
\begin{eqnarray}
N_u = -\begin{pmatrix}
0 & m_c & 0 \\ 0 & 0 & 0 \\ m_t & 0 & 0 \end{pmatrix} ,
\quad
N_d = -\begin{pmatrix}
0 & m_s & 0 \\ 0 & 0 & 0 \\ m_b & 0 & 0 \end{pmatrix}.
\label{e:nund1}
\end{eqnarray}
Clearly, due to small number of parameters in the Yukawa sector, the
FCNC couplings are completely determined in terms of the known physical
parameters. One should keep in mind that \Eqn{e:nund1}
represents the FCNC couplings at the leading order, {\it i.e.},
when the CKM matrix is block-diagonal and the first generation
quark masses are zero. In a more complete theory, these FCNC couplings are
expected to receive small corrections.
But it is still encouraging to note that already at this leading
order, the FCNCs in the down sector are suppressed at least
by $m_b/v$, which means they are quite small in this model. Consequently,
the lower limits on the nonstandard scalar masses are brought down to about 3~TeV as opposed to about 100~TeV for $\order(1)$ FCNC
couplings\cite{Shanker:1981mj,Branco:2011iw}. This makes our model testable at the collider
experiments.

\begin{figure}
\centering
\includegraphics[scale=0.4]{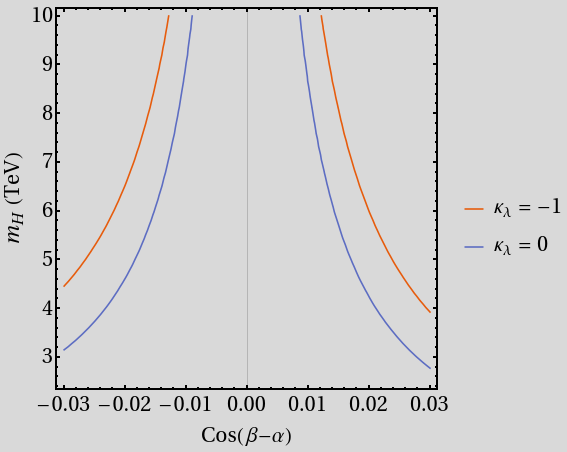}
\caption{\it Contours of $\kappa_\lambda$ in the $\cos(\beta-\alpha)$-$m_H$
    plane. In drawing this plot, we have assumed $\sin2\beta=0.22$ and
    $m_h=125$~GeV.}
\label{f:klam}
\end{figure}

A more interesting scenario arises if one considers slight departure
from the exact alignment limit by turning on small values of
$\cos(\beta-\alpha)$. But one should keep in mind that FCNC couplings
mediated by the SM-like Higgs boson, $h$, will start to seep in via
such a misalignment. However, as mentioned earlier, since the FCNC
couplings are already suppressed, $|\cos(\beta-\alpha)|\lesssim 3\%$
will still be consistent with the flavor data. Such a deviation from
the alignment limit can, in principle, be sensed as tiny deficits in
the Higgs signal strengths because the tree-level couplings of $h$
are suppressed by $\sin(\beta-\alpha)$. Now, we combine this with the
measurement of the trilinear Higgs coupling via Higgs pair production
which probes the following quantity,
\begin{eqnarray}
\kappa_\lambda =\frac{\lambda_{hhh}}{(\lambda_{hhh})^{\rm SM}} =
\frac{\sin(\beta-\alpha)}{2m_h^2\sin4\beta}\left[ 
(m_H^2-m_h^2)\sin4\alpha +(m_H^2+m_h^2)\sin4\beta
+2(m_H^2-m_h^2)\sin2(\alpha+\beta) \right] \,.
\label{e:klam}
\end{eqnarray}
One can easily check that $\kappa_\lambda=1$, as expected in the
limit $\alpha=\beta-\pi/2$. Currently the bound on $\kappa_\lambda$
is quite weak\cite{Sirunyan:2018two,ATLAS-CONF-2018-043}. Note that, the values of $m_h$ and $\beta$,
appearing on the right hand side of \Eqn{e:klam}, are known in our
model. Therefore, assuming that $\cos(\beta-\alpha)$ and $\kappa_\lambda$
settle for some nonstandard values, we can {\em predict} the value of
$m_H$. This feature has been illustrated in Fig.~\ref{f:klam}, where
we can see that if, for example, the values of $\cos(\beta-\alpha)$
and $\kappa_\lambda$ are found to be $0.025$ and $-1$ respectively,
then we can conclude that there should be a heavy neutral scalar
appearing at around $5$~TeV. Although probing such a tiny value of
$\cos(\beta-\alpha)$ might be an ambitious task for the near future,
our model, nevertheless, exemplifies how the Higgs precision
measurements can play a crucial role in pinning down the scale of
new physics.

%%%%%%%%%%%%%%%%%%%%%%%%%%%%%%%%%%%%%%%%%%%%%
%%%%%%%%%%  Conclusions  %%%%%%%%%%%%%%%%%%%%
%%%%%%%%%%%%%%%%%%%%%%%%%%%%%%%%%%%%%%%%%%%%%
To summarize, in this article we have pointed out an intriguing possibility
that there might be a connection between the Cabibbo angle and $\tan\beta$
in a 2HDM. To our knowledge, such a possibility has not been emphasized
earlier in the context of 2HDMs. We accomplish this in a 2HDM with a $D_4$
symmetry which is only softly broken in the scalar potential. Because of
the small number of Yukawa parameters, all the FCNC couplings are completely
determined in our model. Additionally, the FCNC couplings are sufficiently
small so that relatively light scalars accessible at the colliders can successfully
negotiate the flavor constraints. Although the complete CKM matrix and the
exact nonzero masses of the first generation of quarks have not been reproduced
in our minimalistic scenario, we believe that the interesting features of this
model outweigh the dissatisfaction with the small parameters in the quark sector.
Perhaps the present framework can be taken as the first step towards a more
complete theoretical construction which can address the full structure of the
quark masses and mixings. Finally, it should be noted that, although there are quite a few
previous examples of the use of $D_4$ symmetry to understand the leptonic
sector\cite{Adulpravitchai:2008yp,Ishimori:2008gp,Hagedorn:2010mq,Meloni:2011cc,Vien:2014soa,Vien:2013zra}, instances where $D_4$ symmetry has been employed to
explain the quark masses and mixings are rare\cite{Meloni:2011cc,Vien:2019jew} and use four Higgs-doublets. Therefore, the
current paper should be considered as a simpler alternative and an interesting addition to the
existing literature on model building using $D_4$ flavor symmetry.

\paragraph*{Acknowledgments:}
This work has been supported by the Swedish Research Council, contract number 2016-05996.

%%%%%%%%%%%%%%%%%%%%%%%%%%%%%%%%%%%%%%%%%%%%%
%%%%%%%%%%%%%%%%%   References  %%%%%%%%%%%%%
%%%%%%%%%%%%%%%%%%%%%%%%%%%%%%%%%%%%%%%%%%%%%

\bibliographystyle{JHEP} 
\bibliography{Yuk.bib}
\end{document}